\documentstyle[12pt,epsfig,here]{article}
\begin{document}
\newcommand{\beq}{\begin{equation}}
\newcommand{\eeq}{\end{equation}}
\def\beqn{\begin{eqnarray}}
\def\eeqn{\end{eqnarray}}

\newcommand{\Tr}{{\rm Tr}\,}
\newcommand{\E}{{\cal E}}

\newcommand{\ntwo}{${\cal N}=2\;$}
\newcommand{\none}{${\cal N}=1\;$}
\newcommand{\noneh}{${\cal N}=\,
^{\mbox{\small 1}}\!/\mbox{\small 2}\;$}
\newcommand{\vp}{\varphi}
\newcommand{\ve}{\varepsilon}
\newcommand{\pt}{\partial}

\begin{flushright}
FTPI-MINN-06-20\\
UMN-TH-2507-06
\end{flushright}

 \vspace{3mm}

\begin{center}
{\bf {\large\bf Persistent Challenges of Quantum Chromodynamics} \\[1mm]
Julius Edgar Lilienfeld Prize Lecture\\[1mm]
at the April Meeting of APS, Dallas, TX, April 22--25, 2006}

 \vspace{5mm}
 
{\large M. Shifman}

 \vspace{3mm}
 
{\it  William I. Fine Theoretical Physics Institute,
University of Minnesota,
Minneapolis, MN 55455}

 \vspace{3cm}
 
 {\em Abstract}

\end{center}

Unlike some models whose relevance to Nature is still a big question mark,
Quantum Chromodynamics will stay with us forever.
Quantum Chromodynamics (QCD), born in 1973, is a very rich theory  supposed to describe
the widest range of strong interaction phenomena:
from nuclear physics to Regge behavior at large $E$,
from color confinement to quark-gluon matter at high 
densities/temperatures (neutron stars);  the vast horizons of the hadronic world:
chiral dynamics, glueballs, exotics,  light and heavy quarkonia
and mixtures thereof,  
exclusive and inclusive phenomena, interplay between strong forces and weak
interactions, etc. Efforts aimed at  solving the underlying
theory, QCD, continue. In a remarkable entanglement, theoretical constructions of the 1970s and 1990s combine with today's ideas based on holographic description and strong--weak coupling duality, to provide new insights and a deeper understanding.
 
 \vspace{2mm}
 \begin{flushright}
  {\em April 23, 2006}
\end{flushright}

\newpage

 \vspace{1mm}

Unlike some models whose relevance to Nature is still a big question mark,
Quantum Chromodynamics will stay with us forever.
QCD is a very rich theory  supposed to describe
the widest range of strong interaction phenomena:
from nuclear physics to Regge behavior at large $E$,
from color confinement to quark-gluon matter at high 
densities/temperatures (neutron stars);  the vast horizons of the hadronic world:
chiral dynamics, glueballs, exotics,  light and heavy quarkonia
and mixtures  thereof,  
exclusive and inclusive phenomena, interplay between strong forces and weak
interactions, etc. Given the remarkable variety of phenomena governed by QCD dynamics,
it seems unlikely that an exact solution will be ever found.
But do we really need it?

\vspace{-0.1cm}

\begin{center}
 
 {\large\em Birth and Adolescence}

\end{center}

\vspace{-0.1cm}

Quantum Chromodynamics was born in 1973, with the discovery of asymptotic freedom
by David Gross, Frank Wilczek and David Politzer. This discovery was marked by the Nobel Prize in 2004.
 In three decades that elapsed from the beginning
of this exciting journey,
QCD went a long way. Although its full analytic solution has never been found
(and, most likely, never will be), the progress is enormous,
and so are the problems which still await their solutions.
From success to challenge to new discovery --- this is the logic.

I was asked to prepare the Lilienfeld Prize Lecture.
This talk gives me a good opportunity 
to summarize the main elements of a big picture
that emerged after 1973 and outline some promising
problems for the future, as I see it now. 
Rather than aiming at an exhaustive coverage --- which would certainly be
impossible --- 
I will focus on trends drawing them  in ``broad touches." I will cite no original works, referring the reader to selected  books, review papers and lectures
which can fill this gap. 
What came out? Something like   `` A Brief History of Quantum Chromodynamics."
This is not a treatise of an impartial historian.
I am certainly biased and tend to emphasize   those contributions which produced a strong impact on me personally.\footnote{During the APS Meeting presentation I skipped many topics to meet the time constraint; to make up for that I added quite a large number of pictures
which are omitted in the written version.}

To ease my task, I will divide the subject into  three time intervals, covering
three decades --- from 1973 to  '83, from 1984 to '93, and from 1994 to
the  present,
to be referred to as Eras I, II, and III, respectively. The status of QCD by the end of Era II
is summarized in \cite{1},  and its status at the beginning of the new millennium in
\cite{2,2p}. The most recent developments are reviewed in
\cite{caqcd}.

The first triumph that came with the creation of QCD was
understanding those processes where the dominant role belongs to short-distance dynamics,
such as deep inelastic scattering, or the total cross section of the
$e^+e^-$ annihilation. The fact that such processes
could be described, to a good approximation, by the quark-gluon perturbation theory,
was noted by the fathers of QCD. The reason is the famous asymptotic freedom:
the effective quark-gluon coupling becomes weak at short distances.
The boundary between weak coupling and strong coupling lies at
$\Lambda^{-1}$ where $\Lambda$ is a dynamical scale
not seen in the Lagrangian. It occurs through a dimensional transmutation.

The phenomenon of  asymptotic freedom is very counter-intuitive.
Generations of field theory practitioners believed that in any field
theory a probe charge placed in vacuum gets screened
by opposite charges appearing from vacuum fluctuations.
This is intuitively clear. This is certainly the case in quantum electrodynamics (QED).
If so, the effective charge seen by a large-distance observer
falls off with distance, leading to infrared freedom.
Remarkable as it is, in QCD (and non-Abelian gauge theories
at large) it is not screening but rather anti-screening takes place.
The origin of anti-screening is hard to visualize. Perhaps, that's the reason why
the discovery of asymptotic freedom was such a surprise.
 Unlike all ``conventional" theories, in QCD
the effective coupling constant falls off at {\em short} distances --- the opposite of
infrared freedom of QED.\,\footnote{
I should add that QED is logically incomplete because of the Landau zero charge.
It must be viewed as   a part of a larger asymptotically free theory.
At the same time, QCD  is perfect by itself, with a single exception of
the $CP$ problem which   will be discussed below.
} 
 In the early days of QCD people   referred
to this phenomenon as infrared slavery.

Although conceptually similar to that of QED, the
quark-gluon perturbation theory is technically more contrived.
Understanding how to use
perturbation theory when color is permanently confined at large distances,
and quarks and gluons do not appear in the physical spectrum, as well as
adequate techniques, emerged gradually \cite{3}.
Perturbative QCD, or pQCD as it became known later, currently deals with a broad range of issues, from $\Delta T = 1/2$ rule in kaon decays to small-$x$ physics at HERA,
from widths of heavy quarkonia to jet physics. In spite of an advanced age,
this area continues to grow: recently, insights and inspirations from string theory
resulted in an explosive development in the multiparton amplitudes.
I will review this issue later.

In spite of remarkable successes in pQCD, an issue of great practical importance
is not yet solved. In any short-distance-dominated process
there is a stage where quarks and gluons are transformed into hadrons. 
The corresponding dynamics are essentially Minkowskian. Even if theoretical pQCD predictions can be formulated in terms of Euclidean quantities (such as the moments of
the structure functions in deep inelastic scattering), 
the nonperturbative nature of QCD shows up in the form of exponential (in momentum transfers) corrections
which are very difficult to control theoretically. The
situation becomes much worse in those processes in which no Euclidean description is available, for instance, 
in jet physics. Or, if we need to know the structure functions themselves, rather than their moments. In this case pQCD results must be supplemented by corrections which are likely to be oscillating and suppressed by powers of
large energies and momentum transfers, rather than exponentially.

It is these largely unknown corrections that limit the accuracy of theoretical
predictions so that in many instances they lag behind experimentally achieved accuracy.
The problem goes  under the name of {\em quark-hadron duality violation}
\cite{4}. It presents a serious and persistent challenge inherited from Eras I and II,
a stumbling block which is impossible to bypass.

A spontaneously broken axial symmetry in hadronic physics resulting in the occurrence
of the (pseu\-do)-Goldstone bosons was conjectured
in the 1960's, well before the advent of QCD. In fact, in 1957 Marvin 
Goldberger and Sam Treiman
studied the nucleon matrix element of the axial current, including the pion pole at
$t=M_\pi^2$. Assuming the pole dominance they obtained the celebrated
Goldberger--Treiman relation $g_{\pi N\, N} =g_A M_N F_\pi^{-1}$.
In 1960 Nambu identified pions as (pseu\-do)-Goldstone bosons.
A rapid development of the soft-pion technique ensued, allowing one to analyze
a large number of  processes in low-energy hadronic physics. 
The advent of QCD gave new life to all these studies.
It would be fair to say that a macroscopic approach was replaced by a
microscopic one. As an example, let me mention 
a theory-defying enhancement of $\Delta T = 1/2$ amplitudes in $K\to 2\pi, \,\, 3\pi$
decays, observed in the late 1950's. It remained a mystery for years.
Wilson's renormalization group ideas \cite{Wi} applied in QCD, in conjunction
with the lightness of the $u,d$ and $s$ quark masses
\cite{Ga}, led to a discovery of the ``penguin graphs" (Shifman, Vainshtein, Zakharov, 1974) giving rise to $\Delta T = 1/2$
operators with a mixed chiral structure that are indeed strongly enhanced
\cite{Va}.

Needless to say,
without knowledge of underlying dynamics nothing can be said as to
why the axial SU(3)$_{\rm flavor}$ symmetry (for $u$, $d$, $s$ quarks) is spontaneously broken.
Since QCD is {\em the} theory of hadrons,
it should explain this phenomenon. In 1980 Coleman and Witten,
combining the 't Hooft matching condition with the 't Hooft large-$N$ limit
(which I will discuss shortly), proved that the axial symmetry {\em must} be spontaneously broken, indeed. 
At that time, calculation of the order parameter, the quark condensate
$\langle \bar q q\rangle$, was beyond reach. 

During Era I, the soft pion technique \cite{vzpcac,pcac,pcacl}
evolved  into a well-organized system
combining two structural elements: effective low-energy Lagrangians
and chiral perturbation theory. A highlight of this evolution line 
was Witten's discovery in 1983 of the fact that the
chiral Lagrangian supports solitons --- they had been known as Skyrmions ---
which  could be treated quasiclassically in the 't Hooft large-$N$ limit.
Edward Witten demonstrated that the
quasiclassical Skyrmions (collective excitations of the
Goldstone bosons) are in one-to-one correspondence
with baryons of multicolor QCD. This gave rise to the Skyrmion paradigm
\cite{Sk,sk2}, a model of baryons which experienced 
an explosive development in the
beginning of Era II.  Much later, at the end of Era II, it 
was realized that one could use chiral
Lagrangians to describe the interaction of soft pions with hadrons containing a heavy quark \cite{mw}. 

The Skyrme model presents an elegant description of the QCD baryons at 
large $N$. At the same time, it carries a challenge. Assume that we replace conventional massless quarks in the
fundamental representation of SU($N$)$_{\rm color}$ by unconventional quarks, in a different representation of color, e.g. two-index antisymmetric.\footnote{In the actual world $N=3$. At $N=3$ the quark in the fundamental representation is identical to that
in the  two-index antisymmetric representation.}
The pattern of the spontaneous breaking of the chiral symmetry
in this {\em gedanken} case is well-known. The corresponding chiral Lagrangian is not drastically different from that of QCD. It supports Skyrmions too. 
And yet --- in this case there is no apparent match between Skyrmions and baryons
\cite{asv}. Why? A possible way out was suggested by Stefano Bolognesi
just a few days ago.

Now, it is time to dwell on one of the most crucial developments
of Era I --- the invention of the 't Hooft $1/N$ expansion in 1974. It was further extended 
by Witten in 1979 \cite{tho,wi,coleln}. 

Why it is so hard to deal with QCD, and why are new advancements so painfully slow? 
This is due to the fact that in the vast majority of the hadronic problems there is
no apparent expansion parameter. In hard processes $\Lambda/E$ plays the role of such
a parameter. This explains the successes of pQCD. However, the core of the hadronic physics operates with a different set of questions, for instance,
what are the values of the $\rho\pi\pi$ constant, $\omega\phi$ mixing,
$\Sigma$-hyperon magnetic moment? To which extent are the Regge trajectories linear?
Can one calculate their slopes? What can be said about glueballs and why
they are so resilient against experimental  detection? Where are four-quark states and pentaquarks? What is the structure of the newly discovered  charmonium resonances?
This list goes on and on ... In all these cases we do not see any
obvious  expansion parameter.

Gerard 't Hooft\,\footnote{By the way, 't Hooft in English means   {\sl the Head};
isn't it symbolic?}
 came up with a brilliant idea that the number of colors $N$
(in our world $N=3$) can be treated as a large parameter. Consider multicolor QCD
with the gauge group SU$(N)$, instead of SU(3), 
in the limit $N\to \infty$, while the product $\lambda\equiv g^2\, N$ fixed,
where $g^2$ is the gauge coupling constant. This limit is referred to as the 't Hooft limit,
and $\lambda$ as the 't Hooft coupling.
The quarks are assumed to lie in the fundamental representation of SU$(N)$.

A remarkable feature of the 't Hooft $1/N$ expansion is that 
each term of the expansion is in one-to-one correspondence
with topology of the relevant Feynman graphs. The leading order in $1/N$
describes all planar graphs, the next-to-leading order all graphs that can be drawn on a surface with one handle (torus), the next-to-next-to-leading order
requires two handles, and so on. 
Moreover, each extra quark loop is suppressed by $1/N$.
Thus, multicolor QCD (in the 't Hooft limit)
is significantly simpler than QCD {\em per se}. 

Although consideration of planar graphs dramatically reduces the number of graphs,
this is still a vast class of diagrams. Despite numerous attempts, no solution of planar QCD was ever found.

Nevertheless, the $1/N$ expansion proved to be a powerful tool.
At the qualitative level it allowed one to understand 
a variety of regularities inherent to the hadronic world
which seemed rather mysterious for years. 
These regularities are: an infinite number of the meson resonances for 
given $J^{PC}$ and given flavor content;
the Zweig rule (suppression of transitions between the $\bar qq$
pairs of different flavors); a relative smallness of the 
meson widths; the rarity of the four-quark mesons, and so on.

The general picture emerging from the $1/N$ expansion reminds one of the dual
resonance model of the 1960s and early '70s which gave birth to
string theory \cite{gsw}. This parallel, noted already by 't Hooft, 
is no accident. It gave hope
that a string-based description of ``soft" QCD could be found.
We will discuss QCD strings later;  here I would like to note that today
this dream of generations of QCD practitioners no longer seems Utopian,
although, most likely, the equivalence will not be exact.
Jumping ahead of myself, I will add that a version of planar QCD 
has been recently proven to be equivalent to supersymmetric Yang--Mills (SYM) theory
\cite{asv}, with rather nontrivial consequences that ensued immediately.

The $1/N$ expansion as we knew it at that time,
was applicable for qualitative, not quantitative explorations,
with the single exception of the $\eta^\prime$ meson problem, or the puzzle of the
missing ninth (pseudo)-Goldstone boson, of which I will speak later. 
A quantitative (or, at least, a semi-quantitative) method allowing one to address
many questions of hadronic physics from the list presented above 
was invented in 1978. It goes under the name of Shifman--Vainstein--Zakharov (SVZ) sum rules. 
Although the name is quite awkward, the underlying idea is simple and transparent.

The most peculiar features of QCD, such as color confinement
and spontaneous breaking of the chiral symmetry, critical for the 
formation of the hadronic spectrum and basic hadronic characteristics, 
must be reflected in the structure
of the QCD vacuum. Although this structure is contrived,
with luck its salient features could be encoded in a few 
of the ``most important" 
vacuum condensates (I already mentioned one of them, $\langle \bar q q\rangle$;
another is the gluon condensate). If so, one could try to relate
a wealth of the low-energy hadronic parameters to these few condensates
\cite{ms} through the operator product expansion (OPE). 

The notion of
factorization of short and large distances, the central idea of 
OPE,  was borrowed from Ken Wilson.
The focus of Wilson's work was on statistical 
physics, where the program is also known as the block-spin approach.
Surprisingly, 
in high-energy physics of the early-to-mid 
1970s the framework of OPE was essentially
narrowed 
down to perturbation theory. Seemingly, we were the first to adapt 
the general Wilsonian construction to QCD to systematically include 
power-suppressed 
effects, thus bridging the gap between short and large distances.
This ``bridging" did not lose its significance till this day.
I will comment more on that later,  in connection with AdS/QCD.

This
route --- matching between the short distance 
expansion and long distance representation ---
led to remarkable successes. The SVZ method  was tested, 
and proved to be fruitful in analyzing practically 
every static property of all established low-lying hadronic states, both mesons and 
baryons. Needless to say, with just a few vacuum condensates included in the analysis
one cannot expect predictions to be exact, they are bound to
be approximate. However, in many instances 
agreement between theoretical results and experimental data exceeded optimistic
expectations.

As usual there was a cloud on the horizon, a challenge which gave rise to a new development. We discovered that channels with the vacuum quantum numbers
(more exactly, $J^P = 0^\pm$) are drastically different from all others.
In the 1981 paper\,\footnote{This paper was written with V. Novikov.} entitled {\sl Are All Hadrons Alike?},
we observed that these were precisely the channels where the $1/N$ counting
fails too. Indeed, the flavor mixing in the scalar
$\bar q q$ mesons is maximal, and so is mixing with the gluon degrees of freedom.
There is no trace of the Zweig rule. 
In the $0^-$ $\bar q q$ channel Veneziano and Witten predicted  
$M_{\eta^\prime}^2$ to be suppressed by $1/N$, while in actuality
it {\em exceeds} $M^2_\rho$ which does not scale with $N$. The scalar glueball
whose decay width is predicted to be suppressed by $1/N^2$
is in fact much broader than, say, the $\rho$ meson whose 
decay width $\sim 1/N$, etc., etc., etc. On the other hand, in the same paper we noted that in these particular channels the impact of ``direct" instantons (instantons  will be discussed shortly) is the strongest. If in all other cases, by and large, it could
be neglected in the domain of validity of the SVZ sum rules, for the $0^\pm$
quarkonia and glueballs the dominant nonperturbative effect
was obviously correlated with the instantons. In a bid to quantify 
this circumstance, Shuryak; and Diakonov and Petrov
engineered the instanton liquid model \cite{shu}.

Now I have to return to 1975 when Belavin, Polyakov, Schwarz and Tyupkin (BPST)
discovered instantons in non-Abelian Yang--Mills theories, only two years after the advent of QCD. Originally Sasha Polyakov hoped that instantons could solve
the problem of confinement. Although it did not happen that way
(at, least, not in four dimensions)
the conceptual impact of instantons was radical. First of all,
they revealed a nontrivial vacuum structure in non-Abelian Yang--Mills theories.
They demonstrated that an infinitely-dimensional
space of fields has one particular direction which is topologically nontrivial;   it is curled up in
a circle.

 Quantum mechanics of  systems living on a circle is peculiar.
As well-known from solid state physics, in such systems one has to introduce a hidden parameter of an angular type, a quasimomentum, which is not determined from the Lagrangian, but, rather,
from the boundary conditions on the Bloch-type wave functions.  In QCD this
parameter is called the $\theta$ angle, or the vacuum angle.
Instantons represent tunneling trajectories (in imaginary time)
winding around the circle. 

The tunneling interpretation and the
necessity of the emergence of the $\theta$ parameter was
suggested by Gribov; Callan, Dashen and Gross; and Jackiw and Rebbi, independently,
shortly after the BPST work.

Instantons are quasiclassical objects. The qualitative insight they provide is difficult to overestimate. However, in the quantitative aspect the  BPST instantons (or, more generally, the so-called instanton gas)  proved to be rather useless in  QCD.
The reason is obvious: QCD --- the real thing --- is governed by strong coupling.
And still, can one make definite predictions regarding
 the $\theta$ dependence of physical quantities in the hadronic world?
 
Quite an exhaustive answer to this question
was given on the basis of QCD low-energy theorems.
Low-energy theorems are familiar to field theorists from the 1950s.
QCD gave rise to new ones, which were found, one by one, in the
late 1970s. Using them as a tool, Witten in 
1980 exposed quite a sophisticated $\theta$ dependence of
the QCD vacuum. Much later, in 1998 (Era III) he significantly advanced
understanding of this
issue, this time  using a string perspective.
On the field theory side, one can apply supersymmetry-based methods, see, for example,
 \cite{mashi}. They are especially fruitful
at large $N$ and fully confirm Witten's conclusions regarding the
intertwined vacuum family and the corresponding  $\theta$ dependence
consisting of $N$ branches.
Unfortunately,   here I have no time to dwell
on this topic. 

The advent of QCD  put 
the theory of hadrons on solid footing.  It brought a new problem
from an  unexpected
side, however.  Before QCD people believed  $CP$ conservation  
to be a natural feature of strong interactions.  Alas, it is lost in QCD if $\theta\neq 0$ and the quark masses
do not vanish (we know they do not). 
At $\theta\neq 0$ the theory breaks $P$ and $T$ symmetries.
From the absence of $CP$ breaking in strong interactions one concludes that
experimentally $\theta < 10^{-9}$. One can hardly think that
this  incredible smallness of $\theta$ is just an accident.
Can one find a reason for it?

For quite some time Polyakov thought that this was not an issue.
No matter what the bare value of 
the vacuum angle $\theta_0$ at the ultraviolet scale is,
it will be screened to zero by the same effects that lead to color confinement at large distances. Polyakov even asked a student of his
to prove this hypothesis.

Well, this dream never came true.
In 1980 we proved\,\footnote{By we I mean Vainshtein, Zakharov and myself.} that the observability of $CP$-odd effects
at  $\theta_0\neq 0$ is in one-to-one
correspondence with the solution of U(1) problem.
Namely, assuming that  $\theta_0$ is completely screened would require
restoring the Goldstone status of $\eta^\prime$.
Since this is impossible on empiric grounds
(this was shown by S. Weinberg in 1974),
$\theta_0$ cannot be screened. We are back to square one.

Peccei and Quinn suggested an
elegant way out ---
a mechanism that would screen $\theta_0$ no matter what. 
A ``vacuum relaxation" and vanishing of the physical $\theta$ term
is automatic in this mechanism.
Almost immediately
Weinberg and Wilczek noted that the idea leads, with necessity, to a
new particle, the axion.

Their magnificent work described a cute, little, almost massless axion
which was good in all respects, except that it
was incompatible with data.  
It was not so difficult to eliminate this shortcoming. In 1980
we introduced a {\em phantom axion},  a version of what is
now called an ``invisible axion" (this 
was done simultaneously and independently  by Jihn Kim \cite{sre}). 
The invisible axion became a
standard feature of the present-day theory. 
There are two versions of invisible axions;
both preserve positive features of the original axion and,
simultaneously, avoid unwanted contradictions.\footnote{
The invisible axion of the second
kind was devised by Dine, Fischler, Srednicki; and Zhitnitsky.}

Above I have mentioned the 
U(1) problem more than once.
It is also referred to as the problem of the missing ninth Goldstone
meson. The problem dates back to pre-QCD years,
when current algebra was one of just a few tools available to
theorists in strong interactions. The essence of the issue is excellently
summarized in Steven Weinberg's talk at the XVII International
 Conference on High Energy Physics
\cite{Weinb}. With three massless quark flavors
one can construct nine (classically conserved)
axial currents --- eight forming a flavor
octet, plus a flavor-singlet current. None of the corresponding symmetries is
realized linearly in nature.
As far as the flavor-octet currents are concerned, the corresponding
symmetries are spontaneously broken. This implies
the emergence of eight (pseudo)Goldstone bosons which are very well known:
$\pi,\,\,\eta$ and $K$. However, there is no 
Goldstone boson that would correspond to the  flavor-singlet current.
A natural candidate, the $\eta^\prime$ meson,
is too heavy to do the job. The question 
of where the ninth Goldstone boson hides was a big mystery. 

In 1975 't Hooft was the first to note that the $G\tilde G$ anomaly
in the divergence of the flavor-singlet current is not harmless;
it pushes out the Goldstone pole from the physical sector of the theory
to an unphysical gauge noninvariant part of the Hilbert space.
Thus, the particle spectrum of QCD  was not supposed to contain the
ninth Goldstone boson in the first place. 
In 1979 Witten and Veneziano made the next step.
Using 't Hooft's $1/N$ expansion
and the chiral anomaly formula
they managed to obtain an expression relating the  $\eta^\prime$ mass
to the topological susceptibility of the vacuum in pure Yang--Mills
theory (without quarks). They found that in 't Hooft's $1/N$ expansion
$m_{\eta^\prime}^2$ scales as $1/N$. Moreover,  later 
the vacuum topological susceptibility
was calculated in the instanton liquid model and on lattices.
A reasonably good agreement with the empiric value of $m_{\eta^\prime}^2$
was obtained.

Now I turn to the most important aspect of QCD   (after the discovery itself) ---
color confinement. The founding fathers of QCD
---  Gross, Wilczek and Politzer --- after   observing the growth
of the gauge coupling constant at large distances, speculated that this growth
might be responsible for the fact that quarks and gluons,
clearly detectable at short distances, never appear as asymptotic states.
All hadrons that are seen in nature are color-singlet combinations
of the quark and gluon fields. 
Experiment as well as computer simulations in lattice QCD show
that if one considers a quark-antiquark pair separated by a distance $L$
the energy of this system grows linearly with $L$.

However, Gross, Wilczek and Politzer could  not suggest a mechanism that
would explain this phenomenon, linear confinement.
Are we aware of any dynamical systems that could model 
or serve as analogs for this phenomenon,  inseparability of constituents?

In the mid-1970s Nambu, 't Hooft and Mandelstam put forward
a hypothesis \cite{Mand} which goes under the name of the dual Mei{ss}ner
effect (why it is dual will become clear shortly). They were inspired by a natural phenomenon
which takes place in superconductors.

What happens to a superconducting sample if it is placed in a magnetic field?
As well-known, a superconductor expels magnetic flux. 
A superconducting medium tolerates
no magnetic field inside. Assume we have two long magnets with 
well-separated plus and minus poles. Or, better still,
we find a couple of magnetic monopoles (of opposite magnetic charges)
somewhere in space and bring them
here to experiment with them. Next, suppose we insert this monopole-antimonopole pair
into a superconducting sample, and place them at a large
distance $L$ from each other. The monopole is the source 
of the magnetic flux, the antimonopole is a sink, and in the empty space the flux
would spread out to create a Coulomb attraction. However,
inside the superconductor the magnetic flux cannot spread out,
since it is expelled from the superconducting medium. The energetically favorable
solution to this problem is as follows: a thin flux tube forms between the monopole
and the antimonopole. Inside this tube, known as the Abrikosov vortex,\footnote{
Sometimes it is also referred to as the Abrikosov--Nielsen--Olesen
flux tube, or the ANO string. Nielsen and Olesen considered this topological defect
in the context of relativistic field theory.}
superconductivity is ruined. The Abrikosov flux tube has a nonvanishing energy per unit length, a string tension. Once the string is formed the
energy needed to separate the monopole and antimonopole grows linearly with $L$. 
In superconductivity, the formation of the Abrikosov tubes carrying a
quantized flux of the magnetic field is called the Meissner effect. 

Unlike QED, the gauge group in QCD is non-Abelian.
The quarks are sources of the chromoelectric field, rather than chromomagnetic.
Thus, color confinement of quarks
 through string formation would require
chromoelectric flux tubes.  This is why  the Nambu--'t Hooft--Mandelstam
conjecture represents the dual Meissner effect.
The Meissner effect assumes condensation of the electric charges
and confinement of magnetic monopoles. The {\em dual} Meissner effect
assumes condensation of ``chromomagnetic" charges and confinement
of ``chromoelectric" objects.
 In the 1970s and 80s the 
conjecture was nothing more than a vague idea, since people had no clue
as to  non-Abelian monopoles and non-Abelian strings. Any quantitative development was out of reach.

\begin{center}
 
{\large\em Through Era II}

\end{center}

\vspace{-0.1cm}

I will sail rather quickly through the 1980s since these
were relatively quiet years for QCD. I will dwell on just a few developments.

So much was said about the construction of consistent OPE in QCD 
because, after this  was done in connection with the SVZ sum rules,
it gained a life of its own! The very same OPE constitutes the basis
of the heavy quark expansions which blossomed in the 1990's
in the framework of the heavy quark theory  
\cite{BUS,mw}.  This is a branch of QCD where 
a direct live feedback from experiment still exists, 
which gives special weight to any advancement  in theoretical 
understanding and accuracy of predictions.

Conceptually the expansion in inverse powers of the  heavy quark masses $m_Q$
is similar to other applications of OPE. Technically,
exploring physics of mesons with open charm/beauty one has to deal 
with a number of peculiarities. The vacuum condensates are replaced
by expectation values of certain local operators
over the heavy meson states. The most important are the kinetic energy and 
chromomagnetic operators which are responsible for corrections
proportional to $m_Q^{-2}$. 

The OPE-based description of heavy hadrons,
such as $B$ mesons,  conceived in the 1980s
was further expanded in the early-to-mid
1990s, with new elements added and fresh findings incorporated.  One such finding
was a  heavy quark symmetry. It is also known as the Isgur--Wise symmetry.
A special case of this symmetry,
manifesting itself in the $b$-to-$c$ transition at zero recoil, 
was worked out earlier by Voloshin and myself.
The Isgur--Wise consideration covers generic kinematics in the
limit $m_Q\to\infty$.  The heavy quark symmetry
 combining both flavor and spin symmetries
acts in the heavy quark sector
which during Era II  became the focus of experimental studies
in which unprecedented accuracy was achieved.
By and large I can say that
 in the 1990s a quantitative theory of decays of $c$
and $b$-flavored hadrons was constructed that successfully matched the
experimental accuracy. Many people contributed to this success. Among others
I would like to mention Georgi; Bigi, Uraltsev and  Vainshtein (I belonged to this group too), Manohar, Wise;  and Voloshin.

Let me single out one of the most elegant results
established in this way in the heavy quark physics: the 
absence of the $1/m_Q$ correction to the inclusive decay widths of 
the  heavy-flavor hadrons.
This theorem (the Bigi--Uraltsev--Vainshtein theorem)
made its way into textbooks,
let alone its practical importance for the precision determination of 
$V_{cb}$ from data.  

The second development 
is a significant advancement of the $1/N$ ideas
in application to baryons. It was noted by Gervais and Sakita (1984)
and then thoroughly developed by Dashen, Jenkins and Manohar  (1993-94, see
\cite{Mano}) that a large number of model-independent relations among 
baryonic amplitudes follow from large-$N$ consistency conditions.
The essence of these relations is as follows:
At $N\to \infty$ and $m_s\to 0$ the SU(6) spin-flavor symmetry that connects the six states $u\uparrow$, $u\downarrow$,
$d\uparrow$, $d\downarrow$, and $s\uparrow$, $s\downarrow$ 
becomes exact, 
 implying mass and width degeneracies
among baryons of various quantum numbers, as well as 
relations for magnetic moments, axial couplings,  
and so on. Corrections to this limit 
can be systematically treated by combining $1/N$ and $m_s$
expansions.

Next, we witnessed a gradual development --- spanning at least a decade --- of 
the instanton liquid model which grew into
a consistent many-body (four-dimensional)  problem that was 
solved numerically by Shuryak and collaborators.
One may view it as a summation of fermion interactions
to all orders in the 't Hooft instanton-induced vertex. 

The fourth development to be mentioned here proved to be influential, in hindsight,
although it was not perceived as such at the time.
I mean  
the inception of supersymmetry-based methods in gauge theories at strong 
coupling \cite{shiv}.
The inception of these ideas can be traced back to 1983,
when the exact $\beta$ function (the so-called Novikov--Shifman--Vainshtein--Zakharov, or NSVZ
 $\beta$ function) was found in supersymmetric gluodynamics, and
the first exact calculation of the gluino condensate was carried out. The basic ingredient of the above work was the use of holomorphy in the chiral sector
of the SUSY gauge theories. In 1984 Affleck, Dine and Seiberg
added light matter fields and came up with a beautiful superpotential 
emerging in supersymmetric theories with $N_f=N-1$ which bears their name. (Subsequent numerous results
of this group were focused mainly on the issue of spontaneous SUSY breaking.
This topic lies beyond the scope of the present article.)

The issue of whether the 1983 exact result for the gluino condensate was
also correct continued to preoccupy Arkady Vainshtein and me.
In 1987 we engineered a strategy main elements of which
could be considered as precursors of the advanced-to-perfection  
Seiberg and Seiberg--Witten programs.
Although our final target was strongly coupled supersymmetric gluodynamics,
we deformed the theory by introducing additional matter with a small mass term $m$,
in such a way as to guarantee full Higgsing of the theory. Then it became  
weakly coupled. Building on the Affleck--Dine--Seiberg 
superpotential we exactly calculated the gluino condensate at weak coupling,
where each and every step is under theoretical control. 
We then used the holomorphic dependence of the gluino condensate on the
mass parameter to analytically continue to $m\to\infty$,
where the original supersymmetric gluodynamics is 
recovered.\footnote{{\em En route}, the so-called 4/5 problem  surfaced, which is not solved till today.}

By itself, this was a modest result.
It is not the gluino condensate itself, but, rather, the  emerging methods
of SUSY-based analyses that had serious implications in the 1990s.

Concluding this part I would like to mention Seiberg's 1988 calculation of the leading 
nonperturbative correction in the prepotential of ${\cal N}=2$
SUSY Yang--Mills theory ---  apparently,  a starting point of a journey
which culminated in 1994 when Seiberg and Witten found
their celebrated solution for ${\cal N}=2$ theories. 

\vspace{1cm}

\begin{center}
 
{\large\em Maturity}

\end{center}

\vspace{-0.1cm}

It would be fair to say that Era III started with Seiberg--Witten's
breakthrough. String theorists seemingly adore the word {\sl revolution},
at least with regards to their own discipline.
I do not like it because in real life revolutions never solve problems; instead,
they only bring suffering. That's why, in characterizing the Seiberg--Witten 
construction  
and its consequences, the most appropriate
phrase that comes to my mind is {\sl a long-awaited breakthrough}. 
They considered SU(2) super-Yang--Mills theory with extended
supersymmetry, ${\cal N}=2$. Extended SUSY is even more powerful than 
the minimal one. Basing on holomorphy, analytic properties
following from extended
supersymmetry and continuation from weak to strong coupling,
Seiberg and Witten essentially solved the theory 
in the chiral sector at low energies. They proved that SU(2) is spontaneously broken down to U(1) everywhere on the moduli space.
Thus, the magnetic monopoles and dyons are supported everywhere on the moduli space; at certain points they become massless. Deforming the theory
by introducing a small mass term breaking ${\cal N}=2$
to ${\cal N}=1$ Seiberg and Witten forced the monopoles (dyons)
to condense at these points, triggering the dual Meissner effect.
This was the first honest-to-God demonstration ever that the dual 
Meissner effect can indeed take place in non-Abelian gauge theories.

Shortly after, in 1998, Hanany, Strassler and Zaffaroni discussed
formation and structure of the electric flux tubes in the Seiberg-Witten
model which, being stretched between probe charges, confine them. 
Linear confinement in four-dimensional non-Abelian  theory became a reality!

At this time, euphoria of the first  breakthrough years gave place to a 
more sober attitude.
A more careful examination showed that details of
the Seiberg--Witten confinement are quite different from 
those we expect in QCD-like theories. This is due to the fact that in the 
Seiberg--Witten solution the SU($N$)
gauge symmetry is spontaneously broken in two steps.
At a high scale SU($N$) is broken  down to U(1)$^{N-1}$.
Then complete breaking occurs at a much lower scale,
where the monopoles (dyons) condense. Correspondingly,
the confining strings in the Seiberg--Witten
model are, in fact, the Abelian strings  
of the Abrikosov--Nielsen--Olesen
 type. This  results in a ``wrong"  confinement;
 the ``hadronic'' spectrum  in the Seiberg--Witten
model is much richer than that in QCD.

Only recently people started getting ideas about non-Abelian strings,
so far mostly at weak coupling (for reviews see \cite{To,Eto}). Hanany and Tong;
and Auzzi, Bolognesi, Evslin, Konishi and Yung found in 2003 that such strings are supported in certain regimes in ${\cal N}=2$ supersymmetric gauge theories.
Their most crucial feature  is that they have orientational zero modes
associated with rotation of their  color flux inside a non-Abelian
SU($N$). These orientational  modes make 
these strings genuinely non-Abelian. They are supposed to be dual to QCD strings. Shifman and Yung; and Hanany and Tong observed in 2004 
that the above non-Abelian strings
trap non-Abelian magnetic monopoles. In the dual description the trapped magnetic monopoles
should be identified as gluelumps, of which lattice QCD practitioners 
have been speaking
since 1985 \cite{Gr}. A relatively simple {\em weakly coupled} non-Abelian  model was found
which can serve as a laboratory for studying the Meissner effect
in a controllable setting (Gorsky, Shifman, Yung, 2004).

So far I have scarcely mentioned lattice QCD. This is not because I do not value its achievements but, rather, because I believe this is a totally different discipline
than analytic QCD. Over the years lattice practitioners have 
invested ample efforts in numerical studies of QCD strings. 
The very fact of their existence was firmly confirmed.
At the same time, properties of the QCD strings, especially
fine structure, are not easily accessible in numerical simulations. 
I will give just one example, $k$-strings.

The notion of the $k$ strings was introduced largely in the
context of lattice QCD \cite{Gr}. These are the strings that connect
probe color sources with $n$-ality $k$. For instance, if
we use two very heavy quarks sitting on top of each other as the first color source,
and two heavy antiquarks as the second, the string forming between them is
the $2$-string. 

Until the present day the lattice studies did not reveal a fundamental property of 
QCD prescribing the $k$-string tensions to depend only on the $n$-ality,
rather than on the particular representation of the probe color sources.
For the above-mentioned $2$-strings one gets symmetric and antisymmetric
representation of color. These representations are not identical, but
the string tension $\sigma_2$ is expected to be the same in both cases. 
This is expected but is not observed. In 2003 Adi Armoni and I revisited
this long-standing problem (reviewed in \cite{mas}). Our consideration was based on $1/N$ expansion
implying a wealth of quasistable, ``wrong" strings. It was
realized that although  the ``wrong" excited strings 
eventually  must decay into the ``right" ones, whose 
string tension depends only on the $n$-ality,  in many instances the lifetimes
of the excited strings
scale with $N$ exponentially,\footnote{The decay occurs through production
of a pair of gluelumps, of which I spoke previously.} as $\exp (N^2)$. 
By and large, this could explain the failure to confirm universality
of the $k$-string tensions in numerical simulations. 

On the theoretical side of explorations
of   the non-Abelian strings at strong coupling,
all we have at the moment are various conjectures. In 1995 Douglas and Shenker
suggested a much debated Sine formula for
the $k$-string tension which replaced a Casimir scaling hypothesis
that prevailed previously. Douglas--Shenker's arguments were based on ${\cal N}=2$
super-Yang--Mills model.  The Sine formula got further support from
MQCD, and, later (in 2003),  from a conjectured relation between the $k$-wall tension
in ${\cal N}=1$ Yang--Mills theory and the string tension \cite{mas}.
Moreover, Armoni and I observed that the Sine formula is consistent with
the general large-$N$ expansion while the Casimir scaling is not ---
a rather obvious circumstance  previously overlooked. 

Summarizing the issue of the QCD strings  I will
just say that although contours
of the future construction  became, perhaps, visible,
a huge challenge remains --- transforming these contours
into a fully controllable quantitative construction. 

Let us  turn now to a recent  development of the 't Hooft large-$N$ ideas at the interface of  supersymmetry and QCD, known as {\sl planar equivalence} \cite{asv}.
 Genesis of planar equivalence can be traced
to string theory. In 1998
Kachru and Silverstein studied  
various orbifolds of $R^6$ within the AdS/CFT correspondence, of which I will speak later. Starting from ${\cal N}=4$, they obtained distinct --- but equivalent in the infinite-$N$ limit  ---
four-dimensional daughter gauge field theories with matter, with varying degree of supersymmetry,  all with vanishing $\beta$ functions.\footnote{This statement is
slightly inaccurate; I do not want to dwell on subtleties.}

The next step was made   by Bershadsky, Johansen and Vafa.  These authors
eventually abandoned  AdS/CFT, and string methods at large.
Analyzing gauge field theories {\em per se} they proved
 that  an infinite set of amplitudes 
  in the orbifold daughters of the parent ${\cal N}=4$ theory
 in the large-$N$ limit coincide 
with those of the parent  theory, 
order by order in the gauge coupling. Thus, 
explicitly different theories have the same planar limit, at least perturbatively.

After a few years of relative oblivion, interest in the
issue of planar equivalence was revived by Strassler in 2001.
In the inspiring paper entitled
{\em On Methods for Extracting Exact Nonperturbative Results in 
Nonsupersymmetric Gauge Theories,} he shifted the emphasis away 
from the search   
for supersymmetric daughters, towards engineering QCD-like daughters.
Unfortunately, the orbifold daughters considered by Strassler proved to be rather useless. However, the idea gained momentum, and in 2003
planar equivalence, both perturbative and nonperturbative,
was demonstrated to be valid for orientifold daughters (Armoni, Shifman, Veneziano).
The orientifold daughter of SUSY gluodynamics is a
nonsupersymmetric Yang--Mills theory with one Dirac fermion in the two-index antisymmetric representation of SU($N$).
At $N=3$ the orientifold daughter identically reduces to
one-flavor QCD! Thus, one-flavor QCD is planar-equivalent to
SUSY gluodynamics. This remarkable circumstance allows one to copy
results of these theories from one to another.
For instance, color confinement of one-flavor QCD to supersymmetric Yang--Mills,
and the exact gluino condensate in the opposite direction. 
This is how the quark condensate was calculated,
for the first time analytically, in one-flavor QCD (Armoni, Shifman, Veneziano, 2003).

Above I mentioned that in the 1980s and 90s 
applications of the 't Hooft $1/N$ expansion proliferated.
Although the  $1/N$  expansion definitely
captures basic regularities of the hadronic world, it seems to  underestimate
the role of quark loops.
Take, for instance,  the quark dependence of the vacuum energy. 
In the 't Hooft limit the vacuum energy density
is obviously independent of the quark mass, since all quark loops
die out. At the same time, an estimate based on QCD low-energy
theorems tells us   that changing
the strange-quark mass from $\sim 150$ MeV to zero 
would roughly double the value of the vacuum energy density.
An alternative orientifold large-$N$ expansion suggested by Armoni, Veneziano and
myself fixes this problem. However, by and large, phenomenological implications of
the orientifold  $1/N$ expansion have not yet been studied. 

Here I'd like to make a brief digression about surprises.
Surprises accidentally  occur even in old disciplines. 
This is what happened with the quark-gluon plasma (QGP) state of matter
conjectured in the 1970s \cite{shu}. For thirty years QGP was expected to be
a simple near-ideal gas. When it was discovered at RHIC,
just above the phase transition it turned out \cite{ESh1,ESh2} to be strongly coupled!
Theorists working in this area compare this event
with a (hypothetical) discovery of a sizable previously unknown island,
a {\em terra incognita}, in the middle of the Atlantic.
Recently a hypothesis was formulated according to which the strongly coupled QGP
is a plasma of both electric and magnetic charges \cite{ESh2}.

In the remainder of this talk I will focus exclusively on  interrelations
between string theory and Yang--Mills field theories. 
Some of them have been already mentioned above.
My task is to complete this outline.

String theory which emerged from dual hadronic models
in the late 1960s and 70s, elevated to the
``theory of everything" in the 1980s and early 90s, when it
experienced  an unprecedented expansion,
seemingly  entered, in the beginning of Era III,  a ``return-to-roots"
stage.  Results and techniques
of string/D-brane theory, being applied to non-Abelian field
theories (both, supersymmetric and non-supersymmetric),
have generated numerous predictions of various degree of relevance
for gauge theories at strong coupling. If the latter are, in a sense, dual to
string/D-brane theory ---  as is generally believed to be the case ---
they must support  domain walls (of the D-brane type).
In addition, string/D-brane theory teaches us that a
fundamental string that starts on a confined quark, can end
on such a  domain wall. These features are interesting not just by 
themselves; one can hope that, being established, they will shed light
on regularities inherent to QCD (and now we know, they do).
The task of finding solutions to ``down-to-earth"
problems of QCD and other gauge theories by using
results and techniques of string/D-brane theory
is currently recognized by many as {\em the} goal of the community. 
On the other hand, one can hope 
that the internal logic of development of string theory
will be  fertilized by insights and hints obtained from field theory.

\begin{center}
 
{\large\em $D$ Branes in Field Theory}

\end{center}

\vspace{-0.1cm}

In 1996 Dvali and I published a paper entitled
{\sl Domain Walls in Strongly Coupled Theories}.
We reanalyzed supersymmetric gluodynamics,
found an anomalous $(1,0)$
central charge in  superalgebra, not seen at the classical level,
and argued that this central charge will be saturated
by domain walls interpolating
between vacua with distinct values of the order parameter,
the gluino condensate $\langle \lambda\lambda\rangle$,
labeling $N$ distinct  vacua of the theory. 
We obtained an exact relation expressing the
wall tension in  terms of the gluino condensate
\cite{mashi,shiv}. Minimal walls interpolate between 
vacua $n$ and $n+1$, while $k$-walls interpolate
between $n$ and $n+k$.
In this paper we also suggested a mechanism for
localizing gauge fields on the wall through bulk confinement. 
Later this mechanism was implemented 
in many models.

In  the  1997 paper {\sl Branes and the Dynamics of QCD},
Witten interpreted the above BPS walls as analogs of D-branes.
This is because their tension scales as $N\sim 1/g_s$ rather than 
$1/g_s^2$ typical of solitonic objects (here $g_s$ is the string constant).
Many promising consequences ensued. One of them was the Acharya--Vafa
derivation of the wall world-volume theory (2001). Using a wrapped $D$-brane
picture and certain dualities they identified the $k$-wall world-volume theory
as 1+2 dimensional U($k$) gauge theory with the field content of
${\cal N}=2$ and the Chern-Simons term at level $N$ breaking ${\cal N}=2$ down to
${\cal N}=1$. Later Armoni and Hollowood exploited this set-up to calculate the 
wall-wall binding energy.

Beginning from 2002 Al\"esha Yung and I developed a benchmark
${\cal N}=2$ model, weakly coupled in the bulk (and, thus, fully controllable),
which supports both BPS walls and BPS flux tubes.
We demonstrated that a gauge field is indeed localized on the
wall; for the minimal wall this is a U(1) field while for nonminimal walls
the localized gauge field is non-Abelian. We also found a BPS wall-string junction
related to the gauge field localization. The field-theory string
does end on the BPS wall, after all!
The end-point of the string on the wall, after Polyakov's dualization,
becomes a source of the electric field localized on the wall.
In 2005 Norisuke Sakai and David Tong analyzed generic wall-string configurations.
Following condensed matter physicists they called them {\em boojums}.

Summarizing, we are witnessing a very healthy process of cross-fertilization
between string and field theories.
At first, the relation between string theory
and supersymmetric gauge theories
was mostly a ``one-way street" --- from strings to field theory.
Now it is becoming exceedingly more evident
that field-theoretic methods and results, in their turn,
provide insights in string theory.


\begin{center}
 
{\large\em Multiparton amplitudes}

\end{center}

\vspace{-0.1cm}

We already know that Era I was the triumph of perturbative QCD. 
At the same time,  obtaining  high orders in the perturbative expansion
needed for evaluation of the multiparton scattering amplitudes
was an immense technical challenge. To understand the scale of the 
problem suffice it to have a look
at a single color factor in the five-gluon tree amplitude in terms
of dot products of momentum and polarization vectors, see Fig. 1 in 
\cite{bern}. Due to the gauge nature
of interactions in QCD, the final expressions for the multiparton scattering amplitudes
are orders of magnitude simpler than intermediate expressions.

In 1986 Parke and Taylor proposed a closed formula for the scattering
process of the type ``two gluons of negative helicity
$\longrightarrow (n-2)$ gluons of positive helicity," 
where $n$ is arbitrary. This is called the maximal helicity violating (MHV) amplitude. 
Using off-shell recursion relations Berends and Giele then provided a
proof of the Parke-Taylor proposal.
In the 1990's Bern, Dixon and Kosower pioneered applying string methods
to obtain loop amplitudes in  supersymmetric theories and pure
Yang--Mills. The observed simplicity of these results
led to an even more powerful approach based on unitarity.
Their work resulted in an advanced helicity formalism exhibiting a feature 
of the amplitudes, not apparent from the Feynman rules, 
an astonishing simplicity.
In 2003 Witten uncovered a hidden and elegant mathematical 
structure in terms of algebraic curves in terms of twistor 
variables
in gluon scattering amplitudes: he argued that the 
unexpected simplicity could be understood 
in terms of twistor string theory.\,\footnote{A precursor 
of this, for the special case of MHV amplitudes, was given 
by Nair fifteen years earlier.} This observation created a diverse and thriving community of theorists advancing towards full calculation
of multiparton amplitudes at  tree level and beyond, as it became clear
that loop diagrams in gauge theories have their own hidden symmetry structure.
Most of these results do not directly rely on twistors and twistor string theory,
except for some crucial inspiration. So far, there is no good name for this subject.
Marcus Spradlin noted that an unusually large fraction
of contributors' names
start with the letter B\,.\footnote{E.g. Badger, Bedford, Berger, Bern, Bidder, 
Bjerrum-Bohr,
Brandhuber, Britto, Buchbinder, ... (Of course, one should not forget
about Cachazo, Dixon,  Feng, Forde, Khoze, Kosower, Roiban,  Spradlin, Svr\v{c}ek, Travaglini,  Vaman,  Volovich,  ...). This reminds me of a joke of a proof
given by a physicist that almost all numbers are prime:
one is prime, two is prime, three is prime, five is prime,
while four is an exception just supporting the general rule.
} Therefore, perhaps, we should call it $B$ theory, with B standing for beautiful,
much in the same way as M in $M$ theory stands for magic.
I could mention  a third reason for ``$B$~theory":
 Witten linked the scattering amplitudes to a
topological string known as the ``$B$~model."

$B$ theory revived, at a new level, many methods of the pre-QCD era,
when S-matrix ideas ruled the world. For instance, in a powerful paper due to Britto, Cachazo, Feng and Witten (2005), tree-level on-shell amplitudes were shown in a 
very simple and general way to obey recursion relations.
   
Returning to topological string theory in twistor space let me note that it is
dual to a weakly coupled ${\cal N}=4$ gauge theory. Evaluation of the 
string-theory instanton contributions gave  MHV
scattering amplitudes for an arbitrary number of partons   
(Cachazo, Svr\v{c}ek, Witten).  Other amplitudes were
presented  as integrals over the moduli space of holomorphic curves in 
the twistor space (Roiban, Spradlin, Volovich). In essence, the formalism
that came into being   in this way reduces calculations of gauge amplitudes to an effective scalar perturbation theory.
Currently the boundaries of the explored territory are expanding into
loop amplitudes, and there is even a proposal for an all-loop-order 
resummation of MHV planar amplitudes.
A suspicion emerged that planar ${\cal N}=4$ gauge theory
may prove to be integrable!
For  reviews see \cite{kho,fcps}. 

\begin{center}
 
{\large\em Spin Chains and Integrability}

\end{center}

\vspace{-0.1cm}

QCD practitioners ``observed experimentally" rather long ago
that a hidden integrability unexpectedly shows up in problems associated with certain limits of QCD, e.g. high energy Regge behavior of  scattering
amplitudes (Lipatov; Faddeev and Korchemsky, 1994) and the 
spectrum of anomalous dimensions of operators
appearing in deep inelastic scattering (Braun, Derkachov and Manashov, 
1998).
It turns out that in both 
cases evolution equations (in the logarithm
of the appropriate energy scale) can be be identified with  
time evolution governed by Hamiltonians of various integrable quantum spin chains, 
generalizations of  the Heisenberg spin magnet \cite {Beli}.
Historically, integrability was first discovered 
in the Regge limit of QCD. Its relation to evolution equations
for maximal-helicity operators remains unclear till today.

In pure Yang--Mills theories integrability is firmly established to be a property
of one- and two-loop planar evolution equations.\,\footnote{So far
nobody knows what exactly happens beyond two loops.}
At finite $N$ nonplanar corrections  break it.
One can hardly doubt that  integrability  is a consequence
of a general hidden symmetry of all Yang--Mills theories in the limit $N\to\infty$,
not seen at the classical level; it appears dynamically at the quantum level. 
What is its origin? That's where, as many believe, insights from string theory
could help.

A few years later, in 2002-03,
Minahan and Zarembo; and Beisert, Kristjansen and Staudacher
rediscovered the same phenomenon from a different side. These theorists,
motivated by gauge-string duality  (which will be discussed shortly),
studied \cite{beis} renormalization of composite operators in
the maximally supersymmetric field theory,
${\cal N}=4$.

Note that the spectrum of the anomalous dimensions of appropriately chosen operators
is ideally suited for mapping. The dilatation operator on the field-theory side is
identified with a Hamiltonian on the string-theory side. Then anomalous dimensions
of the operators under considerations are mapped onto the energy of the
corresponding string configurations (Berenstein,  Maldacena, Nastase;
and Gubser, Klebanov and Polyakov, 2002, see \cite{ts}). There are 
independent reasons to expect integrability on the string side too 
(Bena, Polchinski and Roiban, 2003). If so, it should 
be valid both for small and large values of the 't Hooft coupling.

This is an excellent example of how QCD and string theory work hand in hand, and, being combined, give rise to applications going far beyond
these two theories. Indeed, before this development the studies of
the Heisenberg magnets in solid state physics were limited to spins in the
finite-dimensional (compact) representations of SU(2).
 (The original Heisenberg model solved through the Bethe ansatz 
was built of  the Pauli 
matrices). In the context of the operator spectrum problem one 
encounters a novel type of Heisenberg magnets, with spin operators that are generators 
of the (super)conformal group in the underlying gauge theory. The 
corresponding representations are necessarily infinite-dimensional and, as a 
consequence, the corresponding 
Heisenberg magnets turn out to be noncompact. Noncompact spin chains 
that ``descended" from gauge theories, have a number of stunning
features, interesting on their own.

An ongoing fusion of both communities
gives hope that integrability of certain problems of Yang--Mills will be explained by a hidden symmetry of a (non) critical string.

I would like to emphasize that QCD in its entirety is not integrable,
beyond any doubt. Are there broader implications of the
hidden integrability, going beyond the two problems mentioned above?
Is there hope that 
spin chain dynamics will make scattering of, say, 5 or 10 gluons
exactly calculable?

\begin{center}
 
{\large\em AdS/CFT or string-gauge holographic duality}

\end{center}

\vspace{-0.1cm}

Now I turn the page and open a new chapter, which, although not yet fully written,
caused a lot of excitement.
It may or may not become yet another  breakthrough in QCD. We will see ...

It all started in 1998 when Maldacena; Gubser, Klebanov and Polyakov; and Witten
argued (conjectured) that certain four-dimensional super-Yang--Mills theories
at large $N$ could be viewed as holographic images
of higher-dimensional string theory. In  the limit of a large 't Hooft coupling
the latter was shown to reduce to anti-de-Sitter supergravity.
The framework got the name ``Anti-de-Sitter/Conformal Field Theory (AdS/CFT) correspondence."

Duality is not something totally new 
in non-Abelian gauge theories. In fact, the first observation
of the Montonen--Olive duality
dates back to 1977, i.e. Era I. Montonen and Olive suggested that
in  four-dimensional  ${\cal N}=4$ super-Yang--Mills theory,
replacing everything ``electric" by everything ``magnetic," 
one obtains an equivalent theory provided that
simultaneously, $g$ is replaced by $1/g$. This is an example of the electric-magnetic
duality; its  ${\cal N}=2$ cousin played an important role in the Seiberg--Witten 
demonstration of the dual Meissner effect.\,\footnote{
The Montonen--Olive duality is the oldest known example of $S$-duality, or a 
strong-weak duality. I must admit that in 1977 I did not appreciate the importance of this result, since I could not imagine, even in my wildest
dreams, that extended supersymmetry could become relevant to QCD.}
AdS/CFT is a totally different kind of duality --- it is holographic.

By now, it is generally believed that ten-dimensional string theory in suitable space-time backgrounds can have a dual, holographic description in terms of superconformal
gauge field theories in four dimensions. Conceptually, the
 idea of a string-gauge duality ascends  to 't Hooft, who realized that the perturbative expansion of  SU($N$) gauge field theory in the large $N$  limit (with the 't Hooft coupling fixed)
 can be reinterpreted as a genus expansion of discretized two-dimensional surfaces built from the field theory Feynman diagrams. 
This  expansion resembles the string theory perturbative expansion  in the string coupling constant.
The AdS/CFT correspondence is a quantitative realization of this idea for four-dimensional gauge theories. In its purest form it identifies the ``fundamental type IIB superstring in a ten-dimensional anti-de-Sitter  space-time background AdS$_5\times S^5$
with the maximally supersymmetric ${\cal N} =4$ Yang--Mills theory with gauge group SU($N$) in four dimensions." The latter theory is superconformal.

At this point I planned, originally, to make a few explanatory remarks. Fortunately, I realized in time
that this will be just another Zhukovsky anecdote. Let me tell you 
this joke (some say that was a true story). Nikolai Zhukovsky was a famous Russian scientist in the fields of gas and fluid mechanics and aeronautics, the theoretical father of Russian aviation. Aeronautics was a very popular subject at the end of the 19th
century;  Zhukovsky began a trend of lecturing for the general public.
At one of the lectures he delivered at the Polytechnical Museum in Moscow
the audience was mainly composed of middle-aged wives of Russian nobility.
He wanted to explain the Bernoulli law, using as an example a sphere 
in a gas flow. When he said ``sphere" he understood immediately that he lost the audience. So, he patiently explained, ``a sphere is a round object like a ball your children play with." He saw smiles and relief on the faces in the audience, and quickly continued:
``thus, you take this sphere and integrate pressure over its surface..."

I am not going to repeat such a mistake, and will skip explanations, referring
the reader to the review papers \cite{mal,Aha,DiV,DHo}.

The main task is to leave conformality  and get as close to real QCD as possible.
Currently there are two (hopefully convergent) lines of though.
Chronologically the first was the top-down approach 
pioneered by Witten; Polchinski and Strassler; Klebanov and Strassler;
Maldacena and Nu\~nez, and others. Here people try to obtain honest-to-God solutions
of the ten-dimensional equations of motion, often in the limit of
a large 't Hooft coupling when on the string side of the theory one deals with supergravity limit. The problem is: in many instances
these solutions
are dual to ... sort of QCD, rather than QCD as we know it. For instance, 
Witten's set-up or the Maldacena--Nu\~nez solution guarantee color confinement but 
the asymptotically free regime of QCD is not attained.

The Klebanov--Strassler supergravity 
solution  is near AdS$_5$ in the ultraviolet limit, a crucial property for the existence of a dual four-dimensional 
gauge theory. In the ultraviolet this theory exhibits logarithmic running of the couplings which goes under the name of duality cascade. They start from string theory on
a warped deformed conifold and discover a cascade of
SU($kM)\times$SU($(k-1)M$) supersymmetric gauge theories on the other side.
As the theory flows in the infrared, $k$ repeatedly changes by unity, 
see the review paper \cite{mast}. In the infrared
this theory exhibits a dynamical generation of the
scale parameter $\Lambda$, which manifests itself in the deformation of the conifold on the string side.

There is a variant  of the 
top-down approach in which the requirement of
the exact solution of the supergravity equations is
``minimally" relaxed. Confinement is enforced
through a crude cut-off 
of the AdS bulk in the infrared, at $z_0$, where $z$ is the fifth dimension.
 This leads to a ``wrong" confinement.
 In particular, the Regge trajectories do not come out
linear.\,\footnote{A year ago, preparing for a talk, I suddenly realized
that the meson spectrum obtained in this way
identically coincides with the 30-year-old result of Alexander Migdal,
who, sure enough, had no thoughts of supergravity in five dimensions. 
His idea was to approximate logarithms of perturbation theory
by an infinite sum of poles in the ``best possible way."
Then this strategy was abandoned since it contradicts OPE.
Now it has been resurrected in a new incarnation. The reason for the
coincidence of the 1977 and 2005 results is fully clear, see Erlich et al., 2006.
} Keeping the full-blown string theory but still adhering
to the above hard-wall approximation one restores asymptotic linearity of
the Regge trajectories at large angular momenta $J$ or excitation numbers $n$.
In this limit one can then calculate, say, the meson decay rates,
as was done recently by Sonnenschein and collaborators, who
recovered the 1979 Casher--Neuberger--Nussinov (CNN) quasiclassical formula!
Then, I would like to mention the 2006 paper of Brower, Polchinski, Strassler and
Tan entitled {\sl The Pomeron and Gauge-String Duality}.\,\footnote{Lenny Susskind  referred to it in jest
as ``Strassler-this-Strassler-that."} Are we witnessing a come-back of 
the large-scale activities in Pomeron physics whose golden years seemingly ended
with the advent of QCD?

Incorporating fundamental quarks, with the spontaneously broken 
chiral symmetry ($\chi SB$),  is a separate problem which seems to be solved by now.
First, fundamental quarks were introduced, via probe branes,  by Karch and Katz, 
in duals of the Coulomb phase. 
Quarks in confining scenarios were introduced, in the context of the Klebanov--Strassler confining background, through D7 branes, by Sakai and Sonnenschein.
A model that admits a full-blown $\chi SB$  
was developed by Sakai and Sugimoto who embedded D8 (anti)branes
in Witten's set-up. 

By and large, I cannot say that at present AdS/CFT
gives a better (or more insightful) description
of the hadronic world, than, say the ``old" SVZ condensate-based method.
Given a rather crude character 
of the hard-wall and similar  approximations,
perhaps, today one may hope to  extract only  universal information 
on hadronic dynamics, steering clear of all details.

\begin{center}
 
{\large\em From AdS/CFT to AdS/QCD}

\end{center}

\vspace{-0.1cm}

This assessment, shared by many, gave rise to an alternative movement,
which goes under the name of AdS/QCD. This bottom-up approach was pioneered by
Son and Stephanov who were motivated, initially, by an observation made by
Bando, Kugo and Yamawaki.
The starting point of  AdS/QCD is a ``marriage" between the holographic representation
and OPE-based methods,
plus $\chi SB$, plus all other ideas that were developed during Era I.
The strategy is as follows: instead of solving the ten-dimensional theory,
the theorist is supposed to make various conjectures 
in order to {\em guess} 
an appropriate five-dimensional metric encoding as much information on real QCD
as possible, and then, with this metric in hands,
get new insights and make new predictions. In this direction I personally was most impressed by works of 
Joshua Erlich, Andreas Karch, Emanuel Katz, Dam  Son, and Misha Stephanov  
in which  
many features of the SVZ expansions were recovered
(as well as the linearity of the Regge trajectories) from a {\em  very simple ansatz}
for the scalar factor in the five-dimensional metric. It remains to be seen how far this road will lead us. Concluding,  I'd like to suggest for consideration of the proponents a couple of trial questions:

\vspace{1mm}

 $\star$ Find the mass splitting of high radial excitations in the
chiral pairs \\\rule{8mm}{0mm}  (e.g. $\rho_n$ and $A_{1n}$);

\vspace{1mm}

 $\star$ Find the next-to-leading term in the $1/n$ expansion of the width-to-mass
 ratio,
 \\ \rule{8mm}{0mm}  say, $\Gamma (\rho_n) /M(\rho_n)$.
 The leading term is given by the CNN formula.

\vspace{5mm}

\begin{center}
 
 {\large \em Instead of Conclusions}

\end{center}

{ \large\bf By the year 1980:}
 
$\bullet$ OPE-based methods were on the rise;

$\bullet$ some crucial low-energy theorems shedding light on the 
QCD vacuum structure
established;

$\bullet$ dual Meissner effect for color confinement conjectured;

$\bullet$ 
$1/N$ expansion as a useful classification tool suggested;

$\bullet$ SUSY gauge theories constructed and 
studied (almost exclusively, in the perturbative sector);

$\bullet$ instantons/monopoles discovered; 

$\bullet$ hypothesis of the
monopole-particle duality in ${\cal N}=4$ put forward.

This is all. Hints were there, but who could have guessed?

\vspace{3mm}

{\large\bf Now:}

$\bullet$
 OPE-based methods culminated in the 1990's;

$\bullet$ 
$1/N$ expansion became semi-quantitative in some problems;

$\bullet$ 
Triumph of SUSY-based methods for {\em QCD cousins} 
is unquestionable (A significant tool kit developed; the dual Meissner effect in ${\cal N}=2^*$ proven!
Dualities in ${\cal N}=1$ discovered!);

$\bullet$ 
Non-Abelian strings discovered and understood; a large number
of parallels between string theory/D branes and non-Abelian
gauge theories revealed from the field theory side;

$\bullet$ 
AdS/QCD, although still in its infancy, starts bringing fruits;

$\bullet$  
String and QCD practitioners are finally talking to each other, to 
their mutual benefit.

{\large\bf Predictions:} 

\noindent
(indirectly depend on external factors, such as SUSY discovery at LHC, ...) 

\vspace{1mm}

$\star$
SUSY-based methods will proliferate,
allowing one to treat {\em closer} relatives of QCD, as well as important 
aspects of QCD {\em per se};

$\! \! \!\!  \star\star$ These methods will spread to
other strongly-coupled theories, e.g. those
relevant to condensed matter physics;

$\! \! \!\!  \! \! \!\!\!  \star\star\star$
The gap between string theories and {\em realistic}
strong-coupling gauge theories will continue to narrow,
with the two-way exchange of ideas;

$\! \! \!\!  \! \! \!\!\! \! \! \!   \star\star\star\star$
$1/N$  expansion and holographic descriptions of QCD  will grow into  powerful 
{\em quantitative} tools, whose accuracy will be under 
complete theoretical control.

\begin{center}
 
{\large\em Acknowledgments}

\end{center}

\vspace{-0.1cm}

I am deeply grateful to my co-authors, for joy of working together and discussing physics
an infinite number of times, from the very first days of QCD. I would like to thank 
A. Armoni, Z. Bern, V. Braun, G. Korchemsky, A. Ritz, E. Shuryak,  M.~Stephanov,
and A. Tseytlin
for valuable comments on the manuscript.

\end{document}